\documentstyle[12pt,twoside,fleqn,espcrc1]{article}

\def\bp{{\bf p}}

\let\De=\Delta

\let\<=\langle
\let\>=\rangle

\def\dsp{\displaystyle}

\def\beq{\begin{equation}}
\def\eeq{\end{equation}}
\def\ba{\begin{array}}
\def\bea{\begin{eqnarray}}
\def\ea{\end{array}}
\def\eea{\end{eqnarray}}

\def\comment#1{ \hbox{[{\it Comment suppressed here.}\/]} }
\def\hide#1{}

\newcommand{\AmS}{{\protect\the\textfont2
  A\kern-.1667em\lower.5ex\hbox{M}\kern-.125emS}}

\title{
\vspace*{-4cm}
\begin{flushright}
\rm MIT-CTP-2715\\
\rm IASSNS-HEP-98/13\\
\vspace*{2.2cm}
\end{flushright}
Color Superconductivity and Signs of its Formation\thanks{
These ideas were first broached
in discussions (in particular with R. Jaffe) at a November workshop 
at the RIKEN-BNL center. The fact that they are now seeing
the light of day owes much to conversations K.R. had at 
QM97  (in particular with R. Seto.)  We are grateful to
the organizers of QM97 and the RIKEN-BNL workshop, as both were
fruitful. The research of M.A. and F.W.
is supported in part by DOE grant DE-FG02-90ER40542;
that of K.R. is supported in part by DOE cooperative
research agreement DE-FC02-94ER40818.}}
\author{Mark Alford,\address{Institute for Advanced Study, 
Princeton, NJ 08540}
Krishna Rajagopal\address{Massachusetts
	Institute of Technology, Cambridge, MA 02139}
and Frank Wilczek$^a$}

\begin{document}
\maketitle
\thispagestyle{empty}

\begin{abstract}
We study finite density QCD
in an approximation in which the interaction
between quarks is modelled on that induced by instantons.
We sketch the mechanism by which chiral
symmetry restoration at finite density
occurs in this model.
At all densities high enough that the chirally symmetric phase
fills space, we find that color symmetry is broken
by the formation of a $\langle qq \rangle$ 
condensate of quark Cooper pairs.
The formation of this  color superconductor condensate
lowers the energy of the system most if 
the up and down quark chemical potentials are equal.
This suggests that the formation of such a condensate in
a heavy ion collision may be accompanied
by radiation of negative pions, and its decay may yield
more protons than were present in the incident nuclei.
\end{abstract}

\section{Introduction}

In his talk at Quark Matter '97, K.R. described our recent
work on QCD at finite baryon density 
presented in
Ref. \cite{us}, and this paper should
be consulted by those seeking a description of the talk.
Here, we
focus on a speculation concerning 
signatures of the formation of a color superconducting
state in heavy ion collisions.

Asymptotic freedom leads us to expect
that at high density quarks behave nearly freely and form large Fermi 
surfaces, with interactions between the 
quasiparticles at the Fermi surfaces which become weak at
asymptotically high density.
Since the quark-quark interaction is attractive in the color
$\bar 3$ channel, BCS pairing of quarks will occur no matter
how weak the interaction.
Pairs of quarks cannot be color singlets,
and so a $\langle qq \rangle$ condensate inevitably breaks
color symmetry.   This breaking is analogous to the breaking
of electromagnetic gauge invariance in superconductivity.
Color superconductivity implies that five of the eight gluons
are massive and implies that the $U(1)$ gauge boson
which remains massless is a linear combination of the
photon and a gluon.  
Our goal in \cite{us}
was to explore this phase in a context
that is definite, qualitatively reasonable, and yet sufficiently
tractable that likely patterns of symmetry breaking and rough
magnitudes of their effects can be identified 
at non-asymptotic densities where interactions are not weak.


Color superconductivity
requires high densities and low temperatures, so the
most favorable experimental conditions are likely those
at the AGS. 
Conditions at the centers of neutron stars are even
more favorable, but that is not our subject here.
Based on present analyses, it seems unlikely that 
color superconductivity can arise at temperatures
above $100$ MeV.  Therefore, one
must select events from the total data set in an
experiment in a way that focuses on a subset 
of events which happen to
have unusually high density and unusually low
temperature.\cite{seto}  The strategy for doing this would
vary in  different experiments, but one might choose
events with little energy at zero degrees (suggesting
central events in which most of the baryons were stopped
at central rapidity creating high densities) and with
protons at central rapidity with unusually soft transverse
momentum (suggesting that the effective temperature experienced
by the baryons in the dense region was lower than in a typical event.)
Signatures of the formation of a color superconductor phase
can then be sought in the subset of events relative to the entire set.
What, then, should we look for in this sort of an event-by-event
analysis?  We propose one answer to this question in Section 3 below, 
although we expect other answers are also possible.

\section{A Model and Chiral Symmetry Breaking Therein}

We sketch here our variational
treatment of a
two-parameter class of models
having two flavors and three colors of massless quarks.  
The kinetic part of the Hamiltonian is 
that for free quarks, while the
interaction Hamiltonian is a  
four fermion interaction (with coupling $K$) 
which is an idealization
of the  
instanton vertex from QCD. 
In order to mimic the effects of asymptotic freedom,
we multiply the momentum space interaction by
%
a product of form factors each of the form
$F(p)=[\Lambda^2/(p^2 + \Lambda^2)]^\nu$,
one for each of the momenta of the four fermions.
$\Lambda$, of
course, is some effective QCD cutoff scale, which one might anticipate
should be in the range 300 -- 1000 MeV.  $\nu$ parametrizes the shape
of the form factor; we consider $\nu=1/2$ and $\nu=1$.


In \cite{us}, we first consider chiral symmetry
breaking at zero density. 
We choose a variational wave
function which pairs particles and
antiparticles with the same flavor and
color but opposite helicity and opposite
three-momentum.  We derive the gap equation
which determines the chiral gap $\Delta_\chi$ 
as a function of quark number density $n$.
We fix the coupling $K$ for each choice of $\Lambda$ and $\nu$
by requiring $\Delta_\chi=400 {\rm ~MeV}$
at $n=0$.  $\Delta_\chi$ decreases with increasing $n$, 
and vanishes at some $n_c$.
With wave function in hand, we then evaluate the 
energy density and pressure
as functions of $n$.
We find that the 
phase with broken chiral symmetry is unstable at any nonzero density,
with this instability being signalled by negative pressure at all
but the lowest densities.
At $n=n_c$, we switch over to an essentially free quark
phase and the pressure then begins to increase monotonically,
reaching zero at a density $n_0>n_c$ and then becoming positive.
In the presence of
a chiral condensate the negative pressure
associated with increasing vacuum energy
overcompensates the increasing Fermi pressure. 
The
uniform, chiral symmetry broken,
nonzero density phase is mechanically 
unstable and breaks up into 
stable droplets of high
density $n=n_0$ in which the pressure is zero 
and chiral symmetry is restored, surrounded by  
empty space with chiral symmetry broken.  
We identify the droplets of chiral
symmetric phase with physical 
nucleons.  Nothing within the model tells us that
the stable droplets have quark number $3$; nucleons are simply
the only candidates in nature which can be identified with droplets
within which the quark density is nonzero and the chiral condensate
is zero. 
This physical picture has significant
implications for the phase transition, as a function of density, to
restored chiral symmetry.   
The transition should occur by
a mechanism analogous to percolation
as nucleons, seen as pre-formed bags of symmetric phase, merge.  


\section{Color Superconductivity with Unequal Numbers of u and d Quarks}    

We now turn to physics at densities greater than $n_0$,
at which the model describes a uniform phase with no
chiral condensate.
At high density, pairing of particles near the Fermi surface as in the
original BCS scheme becomes more favorable.  
Our Hamiltonian supports
condensation in quark-quark channels.   
The condensation is now between fermions with the same
helicity, and the Hamiltonian selects
antisymmetry in flavor.  
One can therefore have spin 0 --- antisymmetric in spin and therefore
in color, forming a $\bar {\bf 3}$, or spin 1 --- symmetric in spin and
therefore in color, forming a {\bf 6}.   
Here, we only consider the former, because although the spin-symmetric 
color {\bf 6} can arise, the gap in this channel is very small.\cite{us}


In Ref. \cite{us}, we construct a suitable trial wave function in which
the Lorentz scalar
$\langle q^{i\alpha} \,C\gamma^5 q^{j\beta}\varepsilon_{ij}\,
\varepsilon_{\alpha\beta 3}\rangle$ is nonzero. This 
chooses a preferred
direction in color space and breaks color $SU(3) \rightarrow SU(2)$.
Electromagnetism is spontaneously broken but
there is a     linear combination of electric charge
and color hypercharge under which the condensate is neutral, and
which therefore generates an unbroken
$U(1)$ gauge symmetry.  
No flavor symmetries, not even chiral ones, are broken.
It is not difficult to redo the derivation of the
gap equation
satisfied by the superconducting gap parameter
$\Delta$ for the case when
down and up quarks have different chemical potentials: 
$~~\mu_+ = \bar\mu + \delta\mu~~$ for the down quarks
and $~~\mu_- = \bar\mu - \delta\mu~~$ for
the up quarks. One finds 
\bea
1 = \frac{K}{\pi^2} \Biggl\{
\int_{\mu_+}^\infty p^2 dp \frac{F^4(p)} 
{\sqrt{ F^4(p)\De^2 + (p-\bar\mu)^2}}
&+& \int_0^{\mu_-}p^2 dp 
\frac{F^4(p)}{\sqrt{ F^4(p)\De^2 + (\bar\mu-p)^2}}\nonumber\\
&+& \int_0^\infty p^2 dp 
\frac{ F^4(p)}{\sqrt{ F^4(p)\De^2 + (p+\bar\mu)^2}}\Biggr\}\ .
\label{colorgapeq}
\eea
The three terms arise respectively from 
particles above the 
Fermi surface, holes below the Fermi surface,
and antiparticles.  With equal numbers of up and down quarks, 
$\mu_+ = \mu_- = \bar\mu$ and the particle 
and hole integrals diverge logarithmically
at the Fermi surface 
as $\Delta \rightarrow 0$, which signals 
condensation for arbitrarily weak attraction.
However, for $\delta\mu\neq 0$ there is no logarithmic
divergence, and $\Delta$ may vanish.  
For a given $\bar\mu$, as $\delta\mu$ is increased the
domain of integration moves farther and farther from the
logarithmic singularity, and $\Delta$ must decrease in order
for the gap equation to be satisfied.  
Momenta between $\mu_+$ and $\mu_-$ cannot contribute,
since neither particle-particle nor hole-hole
pairing is possible, given that the condensate necessarily
pairs up quarks with down quarks.

As a concrete example, we take 
$\nu=1$ and $\Lambda = 800$ MeV in the form factor, and we work
at a baryon density eight times that in nuclear matter.
This density, which corresponds to a quark number density $n$
of $4.1$ per fm$^3$,
is greater than $n_0$,
and so space is filled by a chirally symmetric color 
superconductor phase with quark number density $n$ 
and energy density $\varepsilon$ given by:
\vfill
\eject
\bea
n &=& \frac{2}{\pi^2}\Biggl\{ 
\int_{\mu_+}^\infty p^2 dp \Biggl(1- \frac{p-\bar\mu}{\sqrt{
F^4(p)\Delta^2+(p-\bar\mu)^2}}\Biggr)
- \int_{0}^{\mu_-} p^2 dp \Biggl(1- \frac{\bar\mu-p}{\sqrt{
F^4(p)\Delta^2+(\bar\mu-p)^2}}\Biggr)\nonumber\\
&-& \int_{0}^{\infty} p^2 dp \Biggl(1- \frac{\bar\mu+p}{\sqrt{
F^4(p)\Delta^2+(\bar\mu+p)^2}}\Biggr)\Biggr\}+\frac{\mu_+^3 + \mu_-^3}
{\pi^2}\ ,\\ 
\varepsilon &=& \frac{2}{\pi^2}\Biggl\{ 
\int_{\mu_+}^\infty p^3 dp \Biggl(1- \frac{p-\bar\mu}{\sqrt{
F^4(p)\Delta^2+(p-\bar\mu)^2}}\Biggr)
- \int_{0}^{\mu_-} p^3 dp \Biggl(1- \frac{\bar\mu-p}{\sqrt{
F^4(p)\Delta^2+(\bar\mu-p)^2}}\Biggr)\nonumber\\
&+&\int_{0}^{\infty} p^3 dp \Biggl(1- \frac{\bar\mu+p}{\sqrt{
F^4(p)\Delta^2+(\bar\mu+p)^2}}\Biggr)\Biggr\}- \frac{\Delta^2}{K} 
+\frac{3(\mu_+^4 + \mu_-^4)}{4\pi^2}\ .
\label{energyeq}
\eea
Working first with $\delta\mu=0$, that is with $\mu_+=\mu_-=\bar\mu$,
we find that $n=4.1$ fm$^{-3}$ corresponds to $\bar\mu =0.534$
and $\Delta=0.156$, both in GeV.  If we create an excess of
down quarks by
increasing $\delta\mu$ while
keeping $\bar\mu$ fixed (this changes $n$) we find that
$\Delta$ decreases, and vanishes when $\delta\mu = .037$.
What is a reasonable value for $\delta\mu$?  In $^{197}{\rm Au}$ there
are 315 down quarks and 276 up quarks.  Keeping the density fixed
but enforcing this down/up ratio requires $\bar\mu=0.535$ and
$\delta\mu=0.012$.  Under these conditions, $\Delta$ is decreased
to $0.128$.  This phase,
with a down/up ratio appropriate for gold nuclei, has an energy
density which exceeds that of the $\delta\mu=0$ phase
with the same density by $(99.8 {\rm ~MeV})^4\sim 13 {\rm ~MeV}/{\rm fm}^3$.
This, then, is the energy density by which the up--down symmetric
phase is favored for these parameters.
This estimate is model dependent.  By varying 
the form factor $F$, it is easy to get a result which is twice
(or half) as large. 

We have demonstrated that the formation of a
color superconducting state lowers the energy 
most if the matter is up--down symmetric.  
In a heavy ion collision, in which down quarks initially
exceed up quarks, as a color superconductor condensate forms
in the densest region near the center of the collision 
it may therefore expel a few down quarks and up antiquarks
(one each per 7.4 fm$^3$ of condensate for the density considered 
above) into the surrounding less dense regions.
These will eventually become negative pions, although
the (inhomogeneous, nonequilibrium) dynamics involved are 
far from simple.  When the
condensate breaks up late in the collision, it likely
yields equal numbers of protons and neutrons. 
The total number of protons in the final state will
therefore be more than twice 79. 
We now answer the question posed
in the introduction.
When experimentalists
compare AGS events selected to have unusually high density
and low temperature with events from an entire data set,
they should look for (i) an increase in the number of protons
per event, and (ii) an increase in the $\pi^-/\pi^+$ ratio.
We close by noting that effects of the kind we have 
sketched here, namely equalization of chemical potentials
due to the formation of a color superconductor in high
density regions,
may prove more dramatic when 
the strange quark is included.  This is work
in progress.


\begin{thebibliography}{9}

\bibitem{us}
M. Alford, K. Rajagopal and F. Wilczek, hep-ph/9711395, Phys.
Lett. {\bf B} to appear. 
See this paper for references.  See also
R. Rapp {\it et al}, hep-ph/9711396.



\bibitem{seto}
R. Seto, private communication.
















\end{thebibliography}
\end{document}